\documentclass[12pt,graphicx]{article}
\usepackage{epsfig}
\begin{document}
\baselineskip 0.6cm

\def\simgt{\mathrel{\lower2.5pt\vbox{\lineskip=0pt\baselineskip=0pt
           \hbox{$>$}\hbox{$\sim$}}}}
\def\simlt{\mathrel{\lower2.5pt\vbox{\lineskip=0pt\baselineskip=0pt
           \hbox{$<$}\hbox{$\sim$}}}}

\begin{titlepage}

\begin{flushright}
UCB-PTH 04/23 \\
LBNL-56209 \\
\end{flushright}

\vskip 2.0cm

\begin{center}

{\Large \bf 
Why Are Neutrinos Light? -- An Alternative$^*$}

\vskip 1.0cm

{\large
Lawrence J.~Hall and Steven J.~Oliver
}

\vskip 0.4cm

{\it Department of Physics, University of California,
                Berkeley, CA 94720} \\
{\it Theoretical Physics Group, Lawrence Berkeley National Laboratory,
                Berkeley, CA 94720}

\vskip 1.2cm

\abstract{We review the recent proposal that neutrinos are light
  because their masses are proportional to a low scale, $f$, of lepton
  flavor symmetry breaking. This mechanism is testable because the
  resulting pseudo-Goldstone bosons, of mass $m_G$, couple strongly
  with the neutrinos, affecting the acoustic oscillations during the eV era 
  of the early universe that generate the peaks in the CMB
  radiation. Characteristic signals result over a very wide range of
  $(f, m_G)$ because of a change in the
  total relativistic energy density and because the neutrinos scatter
  rather than free-stream. Thermodynamics allows a precise calculation
  of the signal, so that observations would not only confirm the 
  late-time neutrino mass mechanism, but could also determine whether the
  neutrino spectrum is degenerate, inverted or hierarchical and
  whether the neutrinos are Dirac or Majorana. 

  The flavor symmetries could also give light sterile states. 
  If the masses of the sterile neutrinos turn on after the MeV era, 
  the LSND oscillations can be explained without
  upsetting big bang nucleosynthesis, and, since the sterile states
  decay to lighter neutrinos and pseudo-Goldstones, without giving too
  much hot dark matter.}

\vspace{1in}

$^*$ Talk given by LJH at the  Fujihara Seminar on Neutrino Mass and Seesaw Mechanism 
held at KEK, Japan, February 2004.

\end{center}
\end{titlepage}

\section{Introduction}

With the confirmation of atmospheric neutrino oscillations in 1998,
and, more recently, of large angle solar oscillations, the burning question of
neutrino physics has become ``Why are the lepton mixing angles so large?''
The more fundamental question, of why the neutrinos have non-zero
masses so much smaller than the charged leptons and quarks, is never
heard --- apparently the answer is already known. Treating the standard
model without right-handed neutrinos as an effective theory below
scale $M$, there is a single interaction appearing at dimension 5:
$\ell \ell hh/M$, where $\ell = (\nu_L,e)$ represents the three lepton
doublets. When the Higgs doublet $h$ acquires a vacuum expectation
value $v$, the neutrinos acquire masses
\begin{equation}  
m_\nu \approx \frac{v^2}{M} \hspace{0.75in} \mbox{from} \hspace{0.75in} 
\frac{1}{M} \ell \ell hh.
\label{eq:d=5}
\end{equation}
The gauge symmetries act on the known particles to guarantee that the
neutrinos will have very small, but non-zero, masses.
However, perhaps the most exciting implication of the atmospheric neutrino data
is that, if $M$ is taken to be the Planck mass, this contribution to the 
neutrino mass is too small to account for the observed oscillation length. 
Neutrino masses are not relics from
the gravitational scale, rather they must be understood    
by a non-gravitational field theory below the Planck scale.

Without doubt, the leading candidate for this new physics is the
see-saw mechanism \cite{seesaw}. Introducing right-handed neutrinos $\nu_R$
with Majorana masses $M_R$ and Yukawa couplings $\lambda \; \ell \nu_R h$, 
the effective theory below $M_R$ contains neutrino masses
\begin{equation}
m_\nu \approx \frac{\lambda^2 v^2}{M_R}
 \hspace{0.75in} \mbox{from} \hspace{0.75in}
\lambda \ell \nu_R h + M_R \nu_R \nu_R.
\label{eq:seesaw}
\end{equation}
The see-saw mechanism is so plausible that we sometimes
forget that we do not know if it is correct. The problem with the
see-saw mechanism is that it is too simple: it does not predict any
particles or interactions at energies below $M_R$, so that it cannot
be directly tested. It is remarkable that the heavy Majorana
right-handed neutrino, with the interaction $\ell \nu_R h$
at high energies, does lead to two indirect tests of the
see-saw.  The cosmological baryon asymmetry may be generated by
leptogenesis \cite{lepto}, and, with the addition of supersymmetry,
lepton flavor violation is generated in the slepton interactions
\cite{LFV}. However, even if reactions such as $\mu
\rightarrow e \gamma$ are observed, these connections alone, while
suggestive, will not convince me that the see-saw is correct. To be
convinced one needs a theory that gives precise numerical predictions
for observables.

It appears to be highly significant that the atmospheric data is
explained if the scale $M_R$ of lepton number breaking is within an
order of magnitude or so of the scale for gauge coupling unification.
This connection with grand unification is very exciting --- within the
context of $SO(10)$ theories, one can aim for very predictive theories
of flavor \cite{DHR}. Precise numerical predictions follow if there are
fewer parameters than observables.
The best hope for the see-saw mechanism appears to be a
highly predictive theory for quark and lepton masses, mixings and CP
violation, including leptogenesis and lepton flavor violation. Several
talks at this meeting suggest that $SO(10)$ is currently the most
promising direction \cite{stuart,jogesh,wilfried}, but so far I am not
convinced by any particular theory.

With the theoretical simplicity of the see-saw mechanism so obvious, it
is with some trepidation that I devote the rest of this talk to an
opposite viewpoint. Perhaps neutrinos are light not because they are
inversely proportional to a large scale of lepton flavor symmetry breaking, $m_\nu
\propto 1/M_R$, but because they are proportional to a low scale of
lepton flavor symmetry breaking $m_\nu \propto f$ \cite{CHOO}. Apparently we
discount this possibility because we think that unknown physics lies
at high energies rather than at low energies. However, neutrino
physics is full of surprises and we should explore all
avenues. Indeed, dark energy suggests that we may have missed other
fundamental physics at low energies.

\section{Late-Time Neutrino Masses}

In the standard model, masses for charged leptons and quarks, $\psi$, 
arise from Yukawa interactions $\lambda \bar{\psi_L} \psi_R h$. 
To understand the wide range of Yukawa couplings one frequently 
constructs theories of flavor at some flavor mass scale, $M_{F_c}$, 
based on flavor symmetries that involve a set of scalar fields, 
$\phi_c$, that are charged under the flavor symmetry. The
subscript c refers to the electrically charged fermion sector. In the effective
theory at energy scales
below $M_{F_c}$, the interactions that generate charged fermion
masses have the form $ (\phi_c/M_{F_c})^{n_c} \; \bar{\psi_L} \psi_R h$.
If the fields $\phi_c$ acquire a hierarchy of vevs, $f_c$, the resulting
charged fermion masses are given by
\begin{equation}
m_c = \left( \frac{f_c}{M_{F_c}} \right)^{n_c} v
 \hspace{0.75in} \mbox{from} \hspace{0.75in}
 \left( \frac{\phi_c}{M_{F_c}} \right)^{n_c} \;
\bar{\psi_L} \psi_R h.
\label{eq:FN}
\end{equation}
The charged lepton and
quark mass hierarchies follow from both the hierarchies in $f_c$ and
from a variety of positive integers $n_c$, determined by the group
theory. This global flavor symmetry breaking leads to a set of
Goldstone bosons, $G_c$ --- familons. The experimental limits from such
flavor-changing processes as $\mu \rightarrow e G_c$ and $K
\rightarrow \pi G_c$ are very strong, and, since the interactions of
$G_c$ are proportional to $1/f_c$, the scale $f_c$ is forced to be larger
than about $10^{11}$ GeV. In these theories, the physics of flavor of
the charged fermions must occur at very high energies.

It is possible to develop a very similar picture to explain the small 
neutrino masses \cite{CHOO}.
For the neutrinos to be naturally much lighter than the charged
fermions, it must be that the above flavor symmetry breaking leaves
the neutrinos massless. The neutrino masses must be protected by some
further approximate flavor symmetry. When this symmetry is broken 
by a set of flavor symmetry breaking vevs, $f$, of fields $\phi$, the
neutrinos finally pick up masses
\begin{equation}
m_{\nu_D} = \left( \frac{f}{M_F} \right)^n v
 \hspace{0.75in} \mbox{from} \hspace{0.75in}
\left( \frac{\phi}{M_F} \right)^n \ell \nu_R h
\label{eq:FND}
\end{equation}
for Dirac neutrinos, or
\begin{equation}
m_{\nu_M} = \left( \frac{f}{M_F} \right)^{2n} \frac{v^2}{M_R}
 \hspace{0.75in} \mbox{from} \hspace{0.75in}
\left( \frac{\phi}{M_F} \right)^n \ell \nu_R h + M_R \nu_R \nu_R
\label{eq:FNM}
\end{equation}
for Majorana neutrinos.
The breaking of the neutrino flavor symmetry
leads to a set of Goldstone bosons, $G$. Actually these
bosons are not exactly massless because the neutrino flavor symmetry
is only approximate, so that $G$ acquire very small masses, $m_G$, and
become pseudo-Goldstone bosons (PGBs). 

A crucial question is how large the symmetry breaking vevs $f$ are ---
there must be some new physics at $f$. 
Since the $G$ interactions are proportional to $1/f$, a low
scale for $f$ will mean that $G$ have large interactions. However,
since the symmetry breaking induced by $f$ leads to mass for only the
neutrinos, not the charged leptons or quarks, $G$ couple only to
neutrinos, and the experimental limits on them are extremely weak. The
strongest limits on $f$ come from cosmology and astrophysics rather
than from the laboratory. The requirement that $G$ not be in thermal
equilibrium during big bang nucleosynthesis gives a limit on $f$ of
approximately
\begin{equation}
f \geq 10 \; \mbox{keV},
\label{eq:flimit}
\end{equation}
and a similar limit results from requiring that $G$ emission did not cool
supernova 1987A too rapidly. These are much stronger than any lab
limit on $\nu_a \rightarrow \nu_b G$. Thus, in contrast to the case 
of quarks and charged leptons,  the physics of neutrino
mass generation need not occur at energies much higher than the weak
scale --- rather the relevant flavor symmetry breaking scale could be 
much lower than the weak scale: GeV, MeV or even in the multi-keV domain. 

In the hot big bang the neutrinos are massless until the phase
transition at which $\phi$ acquire their vevs $f$; hence we call our scheme
``late-time neutrino masses.'' In this scheme the neutrinos are light
because the symmetry breaking scale $f$ is much less than the
fundamental mass scale $M_F$ of the physics of neutrino
masses. There is clearly a very wide range of possibilities for these
mass scales. One simple possibility is for Dirac neutrinos with $M_F
\approx v$ and $n=2$, giving $m_\nu \approx f^2/v$, so that $f \approx
100$ keV. In the Majorana case, small neutrino masses may be
partly due to a small $f/M_F$ and partly due to the usual see-saw
mechanism. However, there is no need for any see-saw --- the
right-handed neutrinos could be at the weak scale. Unlike the see-saw 
mechanism, late-time neutrino masses work equally well for both Dirac 
and Majorana cases. 

The spontaneous breaking of lepton symmetries is hardly new.
However, the triplet Majoron model \cite{tripmajoron}, which was
originally seen as a way of protecting neutrino masses with a
spontaneously broken lepton number symmetry, has long been excluded.
On the other hand, the singlet Majoron model \cite{singmajoron} is viewed as a
possible origin for the heavy right-handed neutrino masses, $M_R$. It is
non-trivial that $f \ll v$ is allowed by data, and this option for
explaining why the neutrinos are so light has been sorely neglected.

A possible objection to late-time neutrino masses is that introducing a new
scale $f$  below the weak scale leads to a new hierarchy
problem. However, it is simple to construct theories where $f$ is
generated and stabilized by a low scale of supersymmetry breaking in
the singlet sector \cite{CHOO,CHOP}, or by having composite
right-handed neutrinos \cite{Okui}.

\section{CMB Signals}

In contrast to the see-saw mechanism, late time neutrino masses introduce new
physics at low energy. In particular, there are very light PGBs with
dimensionless couplings to neutrinos of order $m_\nu / f$.  In the
early universe, successful big bang nucleosynthesis requires that $G$
not be in thermal equilibrium during the MeV era, giving the constraint
$f \geq 10$ keV. One 
might expect that $G$ would therefore be irrelevant to cosmology. 
However, decays and inverse decays, $G \leftrightarrow \nu \nu$ or
$\nu_3 \leftrightarrow \nu_{1,2} G$, depending on $m_G$, and the
scattering process $\nu \nu \rightarrow GG$ have reaction rates with
a recoupling form: as the universe cools, these reaction rates increase
relative to the expansion rate, so that $G$ can enter the thermal
bath after nucleosynthesis. Indeed, any $G$ will become thermalized for
a very wide range of $(f, m_G)$: for $f$ as large as the weak scale
and $m_G$ as large as an MeV, as shown by the size of the signal
region in Figure 1.

Any PGB brought into equilibrium by the eV era will leave a signal in
the cosmic microwave background radiation (CMB). Furthermore, the size
of the signal sheds light on the symmetry structure of neutrino
mass generation, as well as the spectrum of neutrinos and whether they
are Dirac or Majorana. The recoupling process itself does not alter
the total relativistic energy density (usually parameterized by the
effective number of neutrino species $N_\nu$) --- it just shares this
energy density between the neutrinos and the PGBs. It is the processes
happening after recoupling that give the crucial signal, and for a
particular PGB the signal depends on whether its mass $m_G$ is larger
or small than the eV scale. PGBs with $m_G > $ eV decay back to
neutrinos before the eV era. This process occurs at fixed entropy,
meaning that the total relativistic energy density is increased.
The angular peaks of the CMB radiation allow a measurement 
of the total radiation energy density during the eV era:  $N_{\nu_{CMB}}$. 
Our prediction, shown in Table 1, depends on the number of such PGBs
and whether the neutrinos are Dirac or Majorana. This signal is
computed with very simple analytic formulas, similar to the case in
the standard cosmology of $e^+ e^-$ annihilation heating the photons
relative to the neutrinos. The numbers given assume that all $n_G$
PGBs recouple to all three neutrinos before the heaviest start to decay, and the
well-known QED correction of $+0.05$ is not included.

\begin{table}
\begin{center}
\begin{tabular}{|c||c|c|}
\hline
$n_G$ &  Dirac  &  Majorana \\
\hline \hline
    1& 3.09 & 3.18 \\ \hline
    2& 3.18 & 3.34 \\ \hline
    8& 3.62 & 4.08 \\ \hline
    16& 4.08 & -   \\ \hline
\end{tabular}
\caption{The effective number of neutrino species,  $N_{\nu_{CMB}}$,
  during the eV era for 3 Dirac or 3 Majorana neutrinos recoupled to
  $n_G$ PGBs heavier than 1 eV.}
\label{table:N_nu}
\end{center}
\end{table}

The number of PGBs, $n_G$, reflects the original neutrino flavor
symmetry. For 3
Dirac neutrinos, the maximal flavor symmetry is $SU(3)_L \times
SU(3)_R$ leading to 16 PGBs, while for 3 Majorana neutrinos the
maximal possibility is $SU(3)$, leading to 8 PGBs. The first year WMAP
data limited $N_{\nu_{CMB}}$ to the range of about 1 to 6 \cite{WMAP}, and
hence was not sufficiently powerful to see these effects. The sensitivity
expectations on  $N_{\nu_{CMB}}$ from future experiments 
are about  $\pm 0.20$ for Planck and $\pm 0.05$ for CMBPOL.

The signal in  Table 1 assumes that each PGB recouples to all three
species of neutrinos. However, a PGB coupling is proportional to
$m_\nu/f$, where here $m_\nu$ means some entry in the neutrino mass
matrix. For a degenerate spectrum of neutrinos, if a PGB recouples to
one neutrino species it will recouple to all three, since the coupling to 
each neutrino is comparable. On the other hand,
depending on parameters, for a hierarchical spectrum the number of
neutrinos the PGB recouples to, $n_R$, could be 1,2 or 3; and for the
inverted spectrum $n_R =$ 2 or 3, but not 1. The dependence of the
$N_{\nu_{CMB}}$ signal on $n_R$ is illustrated in Table 2.

\begin{table}
\begin{center}
\begin{tabular}{|c||c|c|c|}
\hline
$n_R$           & 1    & 2    & 3    \\ \hline
$N_{\nu_{CMB}}$ & 3.77 & 3.98 & 4.08 \\ \hline
\end{tabular}
\caption{The dependence of the total relativistic energy density during the eV
era,  $N_{\nu_{CMB}}$, on the number of neutrino species, $n_R$, to
which the PGBs recouple. 
The case shown is for 8 PGBs and Majorana neutrinos.}
\label{table:n_R}
\end{center}
\end{table}

A PGB with $m_G <$ eV remains in the bath during the acoustic
oscillations that generate the peaks in the CMB, so there is no signal
in $N_{\nu_{CMB}}$. However, the presence of the PGB in the bath
causes an important change in the physics that generates the
peaks, and hence leads to a new characteristic signal. The standard
calculation of the CMB peaks assumes that the neutrinos do not
scatter during the eV era, rather they free-stream from high
temperature to low temperature regions. This causes a well-known
change in the appearance of the CMB peaks, and can be understood
analytically as a being due to a change in the phase of the acoustic
oscillations \cite{BS}. However, interactions with the
background PGB, whether by $\nu_3 \leftrightarrow \nu_{1,2}
G$ or by $\nu G \leftrightarrow \nu G$, prevent this free-streaming,
and hence remove this phase that occurs in the standard picture. This
gives a clear observational signature: a shift in the CMB peak for
the $n$th multipole of
\begin{equation}
\Delta l_n = 7.8 n_{scatt},
\label{eq:deltal}
\end{equation}
relative to the standard model calculations,  where $n_{scatt}$ is the number of
neutrino species that are prevented
from free-streaming by the interactions with $G$ \cite{CHOO}. Other
physical effects also shift the position of the peaks. But it should
be straightforward to observe this large effect since it is a shift
that is independent of the multipole $n$. Clearly one needs to obtain the position
of as many peaks as possible.

\begin{figure}
\begin{center}
  \includegraphics[width=15cm]{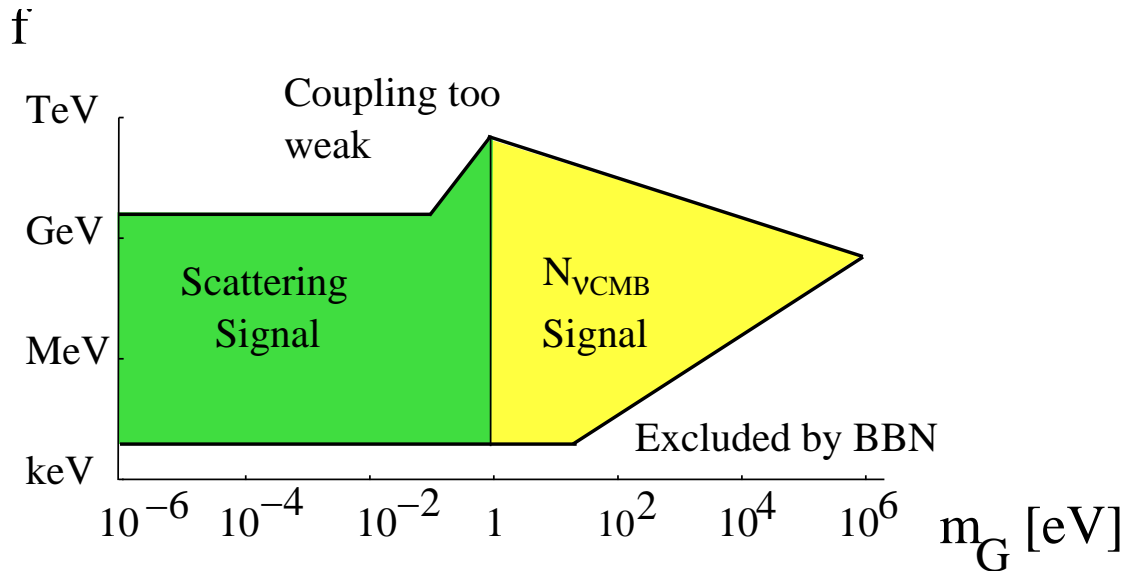}
\caption{Signal regions and cosmological bounds for a
  single Majorana neutrino. CMB
  signals occur throughout the two shaded regions. We have assumed
  $m_\nu = 0.05$ eV.}
 \label{fig:signalregion}
\end{center}
\end{figure}

From the Figure we see that the large signal region in $(f, m_G)$
space is divided into two: an  $N_{\nu_{CMB}}$ signal for $m_G >$ eV
and a $\Delta l_n$ signal for $m_G <$ eV. Since there may be many
PGB, some heavier than others, both signals may be present. We have
taken $m_G$ as a free parameter, since the physical origin of the
explicit flavor symmetry breaking that leads to $m_G$ is very
uncertain \cite{CHOO}. If explicit symmetry breaking arises from dimension 5
operators at the Planck scale, $M_P$, one expects $m_G^2 \approx f^3 /
M_P$, so that signals are expected for $f$ in the range
of 10 keV to 1 GeV.

\section{Lepton Flavor Violation}

In supersymmetric theories, a superpotential interaction of the form 
$\lambda \ell \nu_R h$
will lead to 1-loop radiative corrections in the slepton mass matrix
at order $\lambda^2$, offering the prospect of probing the standard
see-saw mechanism by a variety of lepton flavor violating
interactions \cite{LFV}. However, there is a major limitation to this
signal: the radiative correction occurs at very short distances, governed
by the large right-handed neutrino mass, $M_R$, and is absent if
supersymmetry breaking has not yet appeared as local interactions in
the slepton mass matrix at this high scale. Thus the signal is only present if the  
``messenger'' scale of supersymmetry breaking, $M_{mess}$, is large
enough
\begin{equation}
M_{mess} \geq M_R \simeq v \left( \frac{v}{m_\nu} \right).
\label{eq:mess1}
\end{equation}
Since $M_R$ is typically around $10^{15}$ GeV, this is a very severe
limitation. 

With late-time neutrino masses, the physics of neutrino mass
generation occurs at a much lower scale, enlarging the range of
supersymmetry breaking scenarios for which there will be a lepton
flavor violating signal. At energies below $M_F$ and $M_R$, the
operators responsible for neutrino masses are non-renormalizable, as
shown in (\ref{eq:FND}) and  (\ref{eq:FNM}). To obtain a
sizable signal the radiative correction must involve a renormalizable
interaction. Thus one must go to sufficiently short distances to probe
the origin of these non-renormalizable operators. The radiative
correction can be generated at the scale $M_R$ or $M_F$ ---
much below the scale of $M_R$ in the see-saw mechanism. For example, in
the case of Dirac late-time neutrinos the signal is expected if
\begin{equation}
M_{mess} \geq  f \left( \frac{v}{m_\nu} \right)^{1/n}.
\label{eq:mess2}
\end{equation}
Comparing (\ref{eq:mess1}) with  (\ref{eq:mess2}):
since $f < v$, this opens up a much wider region to lepton flavor
tests, especially for $n>1$.  For example, if $f \approx$ GeV and
$n=2$ the requirement is $M_{mess} \geq 10^7$ GeV.  In the Majorana 
case, signals typically persist for even lower messenger scales.

\section{LSND Neutrinos}

The LSND collaboration reported evidence for the oscillation
$\bar{\nu}_\mu \rightarrow \bar{\nu}_e$ with a probability of $3
\times 10^{-3}$ at the $3.8 \sigma$ level \cite{LSND}. This result is
particularly intriguing: taken with the atmospheric and solar
oscillations, it cannot be explained with the three known neutrinos,
implying the existence of light sterile neutrinos. Combined fits to
neutrino  data are
poor with only one such sterile state, and strongly prefer 2 or more
\cite{SCS} with at least one having a mass above 3 eV.
The theoretical challenge of accounting for the LSND data without
upsetting lab and cosmological constraints is almost insurmountable:
\begin{itemize}
\item Why are there light sterile states? This violates the elegant
  basis for the see-saw mechanism, that fermions without any standard
  model gauge interactions do not have their masses protected and
  hence should be very heavy.
\item Neutrino oscillations in the big bang will populate the sterile
  states before big bang nucleosynthesis, leading to $N_{\nu_{BBN}}
  \geq 5$.
\item Large scale structure surveys and WMAP have put a limit on the
  amount of hot dark matter in the universe, which translates to the
  sum of the neutrino masses being less than 0.7 eV \cite{WMAP+LSS}. 
  The heavy LSND neutrino violates this by at least a factor of 4.
\end{itemize}
A confirmation of the LSND data by MiniBooNE \cite{boone} would herald 
a revolution in neutrino physics more profound than the confirmation 
of atmospheric and solar neutrino oscillations.

Late time neutrino masses provide a perfect reconciliation of the
above difficulties in incorporating LSND oscillations \cite{CHOP}.
\begin{itemize}
\item Lepton flavor symmetries act on right-handed neutrinos as well
  as left, and hence can naturally keep both light.
\item If the flavor symmetries are broken below the MeV scale, $f <$
  MeV, then the neutrinos are exactly massless during the big bang
  nucleosynthesis era. Since only the left-handed neutrinos feel weak
  interactions,  $N_{\nu_{BBN}} = 3$.
\item As the temperature drops beneath the mass of each  sterile
  state, these states are removed from the early universe by decays
  to the three light neutrinos $\nu_s \rightarrow \nu_{1,2,3} G$. 
  The decay rate is guaranteed to be large enough because $f <$ MeV 
  implies large couplings of $G$ to neutrinos, and the observed LSND
  oscillations imply that $\nu_s$ are mixed with the active states.  
  Hence, the limit on hot dark matter applies only
  to the sum of the three light neutrino masses and the PGB
  masses: $m_{\nu_1} + m_{\nu_2} + m_{\nu_3} + \Sigma \; m_G$.
\end{itemize}

The nucleosynthesis constraint implies that $f$ is so low that the
PGBs will necessarily be thermalized before the eV era: if LSND
oscillations are described by late-time neutrino masses, both types of
CMB signal will {\it necessarily} occur. The relativistic energy density signal is now generated
by the decay of the sterile neutrinos  $\nu_s \rightarrow \nu_{1,2,3}
G$, which happens before the eV scale because the sterile states are
heavier than 1 eV. The prediction for  $N_{\nu,{\rm CMB}}$ is shown in Table
3, and depends on the number of PGB, $n_G$, and the number of
sterile neutrinos. 

\begin{table}
\begin{center}
\begin{tabular}{|c||c|c|c||c|c|c|} \hline
    & \multicolumn{3}{|c||}{Dirac} & 
    \multicolumn{3}{|c|}{ Majorana} \\
    $n_G$ & \multicolumn{3}{|c||}{$n_s$} & \multicolumn{3}{|c|}{$n_s$}
    \\
 \hline
     & 1 & 2 & 3 & 1 & 2 & 3 \\ \hline \hline
    2& 3.59 & 3.78 & 3.95 & 3.78 & 3.92 & 4.06  \\ \hline
    3& 3.70 & 3.86 & 4.01 & 3.91 & 4.03 & 4.14  \\ \hline
    8& 4.00 & 4.11 & 4.21 & 4.22 & 4.29 & 4.35  \\ \hline
\end{tabular}
\caption{Effective number of neutrino species during the
recombination era, $N_{\nu,{\rm CMB}}$, in theories with
LSND neutrino oscillations. The signal is produced by $n_s$ 
sterile states decaying to $n_G$ species of PGB and one of the three 
light neutrinos.}
\end{center}
\end{table}

Since the PGBs must be lighter than a few eV in order that
the sterile states can decay, at least some of the light neutrinos
scatter from G exchange, leading to a change in the multipole of the
$n$th CMB peak
\begin{equation}
   \Delta l_n = 23.3 - 13.1 \left( \frac{g_\nu (3-n_{scatt})}{(3g_\nu + n_G)
                (1/N_{\nu, {\rm CMB}}+.23)} \right).
 \label{eq:deltaL}
\end{equation}
The large neutrino mixing angles suggest that all three neutrinos will
scatter, $n_{scatt} = 3$, although $n_{scatt} = 2$ is also possible. The spin
degeneracy is $g_\nu = 7/4,7/2$ for Majorana, Dirac neutrinos. 
In all cases the signal is large.

\section{Conclusions}

While the see-saw explanation of small neutrino masses $m_\nu \propto
1/M_R$ is elegant and plausible, it does not generate any interactions
at low energies that allow it to be directly tested. Even for the
indirect tests, the predictions are only qualitative. We have proposed
an alternative explanation: the neutrino masses are protected by a
small lepton flavor symmetry breaking scale $m_\nu \propto f$
\cite{CHOO}. This symmetry breaking leads to a set of very light
pseudo-Goldstone bosons, of mass $m_G$, with large interactions with
the neutrinos.  These interactions change the cosmological behaviour of
the three known neutrinos before and during the eV era in a way which
leaves a precise and characteristic signal on the CMB radiation. Within
a given model for neutrino mass generation, the consequences for 
the total relativistic energy density and the scattering of the
neutrinos are determined precisely by thermodynamics, and do not
depend on unknown parameters. Hence, observations of these effects
would tell us a great deal about the underlying theory: the symmetry
breaking pattern, the neutrino spectrum and whether the neutrinos are
Majorana or Dirac. Although other physics could lead to a deviation in 
$N_{\nu,{\rm CMB}}$ from 3, in a given theory we are able to make a
precise numerical prediction. Furthermore, a shift in the multipole of
the CMB peaks will tell us that the known neutrinos have a new
interaction at low energy. A combination of both CMB signals, which
results if some PGB are heavier than the eV scale and some lighter,
would be convincing evidence for late-time neutrino masses.

With late time neutrino masses the right-handed neutrinos are much
lighter than in the see-saw case. Hence, in supersymmetric theories
the lepton flavor violating signals are expected for a much wider
range of messenger scales for supersymmetry breaking, enlarging the
interest in this indirect signal. However, it is the precise numerical
predictions, for example of Tables 1,2 and 3 and equations (\ref{eq:deltal})
and  (\ref{eq:deltaL}), that we wish to stress.

The anti-neutrino oscillations observed by the LSND collaboration can
be naturally described by late time neutrino masses; remarkably they
imply signals in both  $N_{\nu,{\rm CMB}}$ and the position of the CMB
peaks \cite{CHOP}. 

\section*{Acknowledgments}

This work was supported in part by the Director, Office of Science, Office 
of High Energy and Nuclear Physics, of the US Department of Energy under 
Contract DE-AC03-76SF00098 and DE-FG03-91ER-40676, and in part by the 
National Science Foundation under grant PHY-00-98840. LJH thanks the 
organizers of the Fujihara Seminar on Neutrino Mass and Seesaw Mechanism 
held at KEK for a stimulating and fruitful meeting.


\begin{thebibliography}{99}

\bibitem{seesaw}
M. Gell-Mann, P. Ramond and R. Slansky, in
$Supergravity$ (North-Holland, Amsterdam, 1979);
T. Yanagida, in {\it Proc. Workshop on Unified Theories and
Baryon Number in the Universe}.

\bibitem{lepto}
M. Fukugita and T. Yanagida,  {\it Phys. Lett.} {\bf 174B} 45 (1986).

\bibitem{LFV}
L.J. Hall, V.A. Kostelecky and S. Raby, {\it Nucl.Phys.} {\bf B267} 415 
(1986);
F. Borzumati and A. Masiero, {\it Phys. Rev. Lett.} {\bf 57} 961 (1986);
J. Hisano, T. Moroi, K. Tobe, M. Yamaguchi and T. Yanagida, 
{\it Phys.Lett.} {\bf B357} (1995) 579, hep-ph/9501407.

\bibitem{DHR}
 S. Dimopoulos, L.J. Hall and Stuart Raby, 
{\it Phys.Rev.Lett.} {\bf 68} 1984 (1992),
{\it Phys.Rev.} {\bf D45} 4192 (1992).
 
\bibitem{stuart}
S. Raby, these proceedings, arXiv:hep-ph/0406022.

\bibitem{jogesh}
J. Pati, these proceedings, arXiv:hep-ph/0407220.

\bibitem{wilfried}
W. Buchmuller, these proceedings, arXiv:hep-ph/0407267.

\bibitem{CHOO}
Z. Chacko, L.J. Hall, T. Okui and S. Oliver, arXiv:hep-ph/0312267.

\bibitem{tripmajoron}
G.B. Gelmini and M. Roncadelli, {\it Phys. Lett.} {\bf 99B}
411 (1981);  H. Georgi, S. Glashow and S. Nussinov, {\it Nucl. Phys.} 
{\bf B193} 297 (1981).

\bibitem{singmajoron}
Y. Chikashige, R.N. Mohapatra and R.D. Peccei, {\it Phys. Lett.} {\bf 98B}
265 (1981).

\bibitem{CHOP}
Z. Chacko, L.J. Hall, S.J. Oliver, M. Perelstein,
arXiv:hep-ph/0405067.

\bibitem{Okui}
T. Okui, arXiv:hep-ph/0405083.

\bibitem{WMAP}
D.~N.~Spergel {\it et al.},
Astrophys.\ J.\ Suppl.\  {\bf 148}, 175 (2003), arXiv:astro-ph/0302209;
P.~de Bernardis {\it et al.}  [Boomerang Collaboration],
Nature {\bf 404}, 955 (2000), arXiv:astro-ph/0004404;
A.~Balbi {\it et al.},
Astrophys.\ J.\  {\bf 545}, L1 (2000)
[Erratum-ibid.\  {\bf 558}, L145 (2001)], arXiv:astro-ph/0005124.

\bibitem{BS}
S.~Bashinsky and U.~Seljak,
{\it Phys.Rev.} {\bf D69} 083002 (2004), arXiv:astro-ph/0310198.

\bibitem{LSND}
LSND Collaboration, Phys.Rev. Lett. {\bf 77} 3082 (1996), nucl-ex/9605003;
Phys.\ Rev.\ C {\bf 58}, 2489 (1998), nucl-ex/9706006;
Phys.\ Rev.\ D {\bf 64}, 112007 (2001), hep-ex/0104049.

\bibitem{SCS}
M.~Sorel, J.~M.~Conrad and M.~Shaevitz,
arXiv:hep-ph/0305255.

\bibitem{WMAP+LSS}
D.~N.~Spergel {\it et al.},
Astrophys.\ J.\ Suppl.\  {\bf 148}, 175 (2003), arXiv:astro-ph/0302209;
W.~J.~Percival {\it et al.},
astro-ph/0105252.

\bibitem{boone}
R.~Tayloe  [MiniBooNE Collaboration],
Nucl.\ Phys.\ Proc.\ Suppl.\  {\bf 118}, 157 (2003).

\end{thebibliography}
\end{document}